\documentclass[12pt]{article}
\usepackage[margin=0.9in]{geometry}

\usepackage[utf8]{inputenc}
\usepackage{bm}
\usepackage{bbm}
\usepackage{stmaryrd}
\usepackage{amsmath}
\usepackage{amsthm}
\usepackage{amssymb}
\usepackage{mathtools}
\usepackage{xcolor}
\usepackage{graphicx}
\usepackage{ulem}
\usepackage{url}
\usepackage{comment}

\renewcommand{\c}{\mathcal C}

\newcommand{\rx}{{\mathbb R}}
\newcommand{\cx}{{\mathbb C}}
\newcommand{\tr}{\mathrm{tr}}

\newcommand{\h}{{\mathcal H}}

\renewcommand{\r}{{\rm R}}

\newcommand{\s}{{\rm S}}

\renewcommand{\i}{{\rm I}}
\newcommand{\vac}{|{\rm vac}\rangle}
\newcommand{\vacbra}{\langle {\rm vac}|}

\newcommand{\bbbone}{\mathchoice {\rm 1\mskip-4mu l} {\rm 1\mskip-4mu l}
{\rm 1\mskip-4.5mu l} {\rm 1\mskip-5mu l}}
\newtheorem{thm}{Theorem}

\newtheorem*{prop*}{Proposition}

\usepackage{cleveref}
\theoremstyle{definition}
\newtheorem{ex}{Exercise}

\usepackage{authblk}

\title{Lectures on Open Quantum Systems}

\author[1]{Marco Merkli\footnote{merkli@mun.ca}}
\author[1]{\'Angel Neira\footnote{aeneira@mun.ca}}

\affil[1]{Department of Mathematics and Statistics

Memorial University of Newfoundland

St.~John’s, NL, Canada A1C 5S7
}

\begin{document}

\maketitle

\newcommand{\ml}{\mathcal L}
\newcommand{\mv}{\mathcal V}

\begin{abstract}
These notes are a short introduction to the mathematical theory of open quantum systems. They are meant to serve as an entry point into a broad research area which has applications across the quantum sciences dealing with systems subjected to external noise. The guiding idea is to let the key structures of the theory emerge from a concrete model. By working through the dissipative Jaynes-Cummings model the reader will discover explicitly how irreversible dynamics arises from a unitary system-reservoir evolution. The notions of the continuous mode limit, correlation functions, spectral density appear in a natural manner and lead to the evolution equation of the open system in form of a  master equation. This sets the stage for the more general analysis of completely positive, trace preserving (CPTP) maps and the study of quantum dynamical semigroups. We motivate and prove the Kraus representation theorem, the dilation theorem and the Gorini-Kossakowski-Sudarshan-Lindblad (GKSL) theorem. Working through the exercises (for which full solutions are supplied) will reinforce the ideas introduced in the main text. 
\end{abstract}

{\bf Guide to the notes.}
The notes are designed to provide a succinct introduction to some aspects of the theory of open quantum systems. They are suitable for  undergraduate and graduate students with little to no prior knowledge of quantum theory. In Chapter \ref{sec:formalism} we present an introduction to the four main postulates of quantum theory: states, dynamics, measurement, composition of subsystems. In Chapter \ref{sec:jaynes} analyze the dissipative Jaynes-Cummings model in detail. Our approach is direct and does not resort to special knowledge about the theory of open quantum systems. The analysis  reveals dissipative (irreversible) dynamical behaviour which is a typical characteristic of open systems. The chapter culminates in the derivation of  the exact (non-Markovian) master equation for the evolution of the dissipative Jaynes-Cummings model. The master equation is generated by a Lindblad (super)operator, which motivates us to explore the structure and properties of such generators in the remainder of the notes. In Chapter \ref{sec:cptpmaps} we analyze completely positive trace preserving (CPTP) maps and we show the famous {\it Kraus Representation Theorem} which describes the structure of all CPTP maps. {\it A priori} CPTP maps are defined by their mathematical properties, but a beautiful result | the {\it Dilation Theorem} | links them to the theory of open quantum systems. Namely, CPTP maps are equivalently expressable as the reduction of a system-plus-reservoir dynamics to the system part alone. We give a proof of this result.  In Chapter \ref{sec:QDSG} we explore {\it quantum dynamical semigroups}, which are mappings satisfying a group composition law, and which associate to an initial density matrix a density matrix at a later time. We derive the famous {\it Gorini-Kossakowski-Sudarshan-Lindblad (GKSL) Theorem} which reveals the general structure of generators of such semigroups. Finally, the notes contain a number of exercises, with solutions provided at the end.

\section{Formalism of quantum theory}
\label{sec:formalism}

The {\bf pure state} of a physical system is a  vector $\psi$ of a Hilbert space $\mathcal{H}$, normalized as $\|\psi\|=1$. Typical examples for systems and their Hilbert state spaces are:
\begin{itemize}    
\item[-] A qubit, or spin $1/2$, or two-level system has the state space  $\mathcal{H}=\mathbb{C}^2$. Basic states are $|0\rangle, |1\rangle$. These kets denote the canonical basis of $\mathbb C^2$. A general state is of the form $\psi=a|0\rangle +b|1\rangle$ with $a,b\in\mathbb C$, $|a|^2+|b|^2=1$.  Similarly, $\mathbb C^N$ is the state space of an $N$-level system.
\item[-] A particle moving in $d$ spatial directions has Hilbert space $\mathcal{H}=L^2(\mathbb{R}^d,d^dx)$, the square-integrable complex valued functions of $x\in\mathbb R^d$. An example of a state is the Gaussian function $\psi(x)= (2\pi\sigma)^{-d/2}e^{-\|x\|^2/2\sigma}$, where $\sigma >0$.
\item[-] The Hilbert space carrying states of a (scalar, bosonic) quantum field ({\it e.g.}~the electromagnetic field) is {\it Fock space} $\mathfrak{F}=\mathbb{C}\bigoplus_{n=1}^{\infty}\mathfrak{h}^{\otimes_{\rm sym} n}$. Here, $\mathfrak h$ is a `single particle Hilbert space', for example $\mathfrak h =L^2(\mathbb R^3,d^3k)$, describing a single particle/excitation. The symmetrized tensor product $\mathfrak{h}^{\otimes_{\rm sym} n}$ is called the $n$-sector. It is the Hilbert space of $n$ (indistinguishable, bosonic) particles. The summand for $n=0$, $\mathbb C$, is called the vacuum sector (absence of particles). The direct sum structure allows the possibility to have a variable number of particles in the system. Operators which connect sectors with different $n$ implement the creation or annihilation of particles. The state $\psi= 1\oplus 0\oplus 0\oplus\cdots\in\mathfrak F$ is the vacuum state. Given $f\in\mathfrak h$ with $\|f\|=1$ (the norm of $\mathfrak h$), the vector $\psi=0\oplus f\oplus 0\oplus\cdots\in\mathfrak F$ is the state of the field in which one particle is present, and this particle is in the state $f\in\mathfrak h$.
\end{itemize}

The {\bf dynamics} of a quantum system is given by a map $t\mapsto \psi_t$, where $\psi_t\in\h$ is a state for all times $t\ge 0$ (or $t\in\mathbb R$) and the `initial condition' is $\psi_0=\psi_{t=0}$. The family $\{\psi_t\}_t$ is a path in the Hilbert space $\h$, called the orbit of $\psi_0$. The dynamical equation is the {\it Schr\"odinger equation}, 
\[
i\hbar \partial_t \psi_t = H \psi_t.
\]
In the sequel we will set Planck's constant $\hbar$ equal to one (which amounts to a rescaling of the time scale or of the energy scale). On the left side we have the imaginary unit $i$ and the time derivative $\partial_t$ and on the right side a self-adjoint ($\equiv$ hermitian) operator $H$ on $\h$, called the {\it Hamiltonian}, acts on the vector $\psi_t$. The Schr\"odinger equation is also written in `integrated form' as
\[
\psi_t =U_t \psi_0\, ,\qquad U_t=e^{-it H}
\]
and the strongly continuous one-parameter unitary group $U_t$ is called the {\it propagator}.

Given a physical system with Hilbert space $\h$, an {\bf observable} is a self-adjoint linear operator on $\mathcal{H}$. Every physically observable quantity corresponds to such an operator. For example, the Hamiltonian $H$ in the Schr\"odinger equation represents the energy of the system. Suppose a {\bf measurement} of an observable $A$ is carried out on a fixed state $\psi$. The measurement outcome is a real number, it is one of the eigenvalues $a$ of the operator $A$ (if $A$ has continuous spectrum then one has to rephrase this somewhat). The measurement outcomes are {\it random}, each outcome $a$ has an associated probability\footnote{In fact, $p_a=\langle \psi, P_a\psi\rangle$, where $P_a$ is the spectral projection of $A$ associated to the eigenvalue $a$ of $A$.} $p_a$. It is a curious fact of quantum theory that a measurement of the {\it same} quantity (namely $A$) on the {\it same} state (namely $\psi$) will yield {\it different} values (eigenvalues $a$) in compliance with a statistics (given by the probabilities $p_a$). One can see that the measurement outcome is `non-random' exactly if the state $\psi$ is an eigenstate of the observable $A$ | then the measurement outcome will be the associated eigenvalue $a$ with $p_a=100\%$ probability. The {\it average}, or {\it expected value} of the measurment readout of the observable $A$ in the state $\psi$ is given by
\[
\langle A\rangle_\psi = \langle\psi, A\psi\rangle.
\]
On the right side, $\langle \cdot, \cdot\rangle$  is the inner product (scalar product) of the Hilbert space. 
\medskip

Some examples of observables and their eigenvectors are
\begin{itemize}
    \item[-] The Pauli operator $\sigma_z=|0\rangle\langle 0|-|1\rangle\langle 1|$, which satisfies
    \[
    \sigma_z|0\rangle=|0\rangle, \quad \sigma_z|1\rangle=-|1\rangle
    \]
    in the standard basis $\{|0\rangle,|1\rangle\}$ of $\mathbb C^2$. After taking a measurement of this observable in any state, the result will be either $1$ or $-1$; it distinguishes  two quantum bits of information.
\item[-] The Hamiltonian of the harmonic oscillator (coordinate representation) is
    \[
    H=-\dfrac{1}{2M}\dfrac{d^2}{dx^2}+\dfrac{M\omega^2}{2}x^2
    \]
and acts on the Hilbert space is $L^2(\mathbb R,dx)$ and has eigenvectors (eigenfunctions) 
    \[
    \psi_n(x)=\Big(\dfrac{M\omega}{\pi2^{2n}(n!)^2}\Big)^{1/4}H_n\Big(\left(M\omega\right)^{1/2}x\Big)\exp\Big\{-\frac{1}{2}M\omega x^2\Big\},\qquad n\ge 0,
    \]
where the $H_n$ are the Hermite polynomials \cite{hermite}. The $\{\psi_n\}_{n\ge 0}$ form an orthonormal basis for $\mathcal H$. The result of a measurement of $H$ in any state is one of the eigenvalues $E_n=\omega(n+1/2)$.  
    
\item[-]  The position operator $X$ of a particle moving on a line acts on $\psi\in\h=L^2(\mathbb R,dx)$ as the multiplication operator,
\[X\psi(x)=x\psi(x).\]
This is an unbounded self-adjoint operator defined on its domain $\{\psi\in L^2(\mathbb R,dx)\, :\, x\psi(x)\in L^2(\mathbb R,dx)\}$. It does not have any eigenvalues, but only  continuous spectrum; therefore, as there are no eigenvectors of $X$, it is impossible to have a particle localized {\it exactly} at a given point in space. Formally, ``$\psi(x)= \delta(x-x_0)$'' (Dirac delta function) would be the state of the particle localized at $x_0$, but this $\psi(x)$ is not square integrable, that is, $\psi\not\in\h$ \cite{Hall2013}. Rather, given any $\psi(x)\in \h$, the quantity $|\psi(x)|^2dx$ is the {\it probability density} for finding the particle at $x$. If the function $|\psi(x)|^2$ is sharply peaked at some $x_0$ then the particle is localized around this position.

The momentum operator $P=-i d/dx$ is, like the position operator, an unbounded self-adjoint operator. It satisfies $Pe^{ipx}=pe^{ipx}$, but again, the functions $e^{ipx}$ are not square integrable, so they don't belong to $\h$. Rather, the Fourier transform yields $|\widehat \psi(p)|^2dp$, the probability density for the measurement of the momentum $p$ of $\psi$.  When dealing with operators with continuous spectra, the theory of generalized eigenvectors is required for a mathematically rigorous formulation \cite{BohmGadella}.

\end{itemize}

The Hilbert space of a {\bf composite system} is the product of the individual Hilbert spaces. For example, a register of $N$ qubits has Hilbert space $\mathcal{H}=\mathbb{C}^2\otimes\mathbb{C}^2\otimes \cdots\otimes\mathbb{C}^2$, the $N$-fold tensor product of $\cx^2$. The Hilbert space of a qubit plus the electromagnetic field (`atom-radiation complex') is $\mathbb{C}^2\otimes\mathfrak{F}$.

Systems which have a natural partition into two parts are called {\it bipartite systems}. An example is the atom and the field above. In this situation each part is called an {\bf open system}, as each part generally exchanges energy, particles, correlations with the other part. Often one is interested in only one of the two parts in a bipartite systems, called `the system' S, while the remaining part is termed the `environment' or the `reservoir' R. The entire complex is then called a $\s\r$ complex. Its Hilbert space is 
\[
\h_{\s\r} = \h_\s\otimes\h_\r.
\] 

The (pure) state of the $\s\r$ complex is given by a normalized vector $\psi\in\h_{\s\r}$. Suppose we are interested in measuring the observable $A$ of the (sub-)system $\s$ in the state $\psi\in\h_{\s\r}$ | $A$ is a linear operator on $\h_\s$. Its expectation is given by $\langle \psi, (A\otimes\bbbone_\r)\psi\rangle$, where the inner product is that of $\h_{\s\r}$. One can express this average purely in terms of quantities pertaining to $\s$ as follows,
\[
\langle \psi, (A\otimes\bbbone_\r)\psi\rangle = {\rm tr}_{\s\r}\Big(|\psi\rangle\langle\psi| (A\otimes\bbbone_\r)\Big)={\rm tr}_\s\Big[\big({\rm tr}_\r |\psi\rangle\langle\psi|\big) A\Big].
\]
Here, ${\rm tr}_\r |\psi\rangle\langle\psi|$ is the {\it partial trace} of $|\psi\rangle\langle\psi|$, taken over the degrees of freedom of $\h_\r$ (see also Section \ref{sec:cptpmaps} for more detail on the partial trace). This is an operator on $\h_\s$, called the {\it reduced density matrix} or the reduced state of the system,
\[
\rho_\s = {\rm tr}_\r |\psi\rangle\langle\psi|.
\]
One readily sees that $\rho_\s$ is a non-negative operator, $\rho_\s=\rho_\s^\dag\ge 0$ (the star denotes the adjoint operator) and it is normalized as ${\rm tr}_\s\, \rho_\s=1$. Generally $\rho_\s$ cannot be written in the form $|\psi_\s\rangle\langle\psi_\s|$. In fact, it turns out that $\rho_\s$ is of this form exactly if the original state is disentangled, that is, if it is of the form $\psi=\psi_\s\otimes\psi_\r$. This means that we should generalize our notion of state of a quantum system: The state of a system with Hilbert space $\h$ is given by an operator $\rho\ge 0$ satisfying ${\rm tr}\rho=1$. Such non-negative, normalized operators are called  {\it density matrices}. According to the above discussion, the expected value of the measurement outcome of an observable $A$ in the state $\rho$ is then
\[
\langle A\rangle_\rho = {\rm tr}\big(\rho A\big).
\]
If $\rho$ has rank one, then $\rho=|\psi\rangle\langle\psi|$ for some normalized $\psi\in\h$ and we recover our initial definitions. In this case $\rho$ is called a {\it pure state}. If the rank of $\rho$ exceeds one then $\rho$ is called a {\it mixed state}.

Consider the dynamics of a pure state of a bipartite $\s\r$ system, $\psi_t=e^{-it H}\psi_0\in\h_{\s\r}$, where $H$ is a Hamiltonian acting on $\h_{\r\s}$. The corresponding reduced system state becomes time dependent as well,
\[
\rho_\s(t) = {\rm tr}_\r\big(|\psi_t\rangle\langle\psi_t| \big) = \tr_\r\left(e^{-itH}|\psi_0\rangle\langle\psi_0|e^{itH}\right).
\]
Note that the factor $e^{itH}$ cannot be moved to the left of $e^{-itH}$ in this expression because $\tr_\r$ is the {\it partial} trace only, which is cyclic for operators of the form $\bbbone_\s\otimes A_\r$ only. The reduced density matrix $\rho_\s(t)$ is never of the form $e^{-itH_{\rm eff}}\rho_\s(0)e^{itH_{\rm eff}}$ for some `effective Hamiltonian' $H_{\rm eff}$ acting on $\h_\s$ | except in the uninteresting case when the system and reservoir are not interacting. In fact, the map $t\mapsto \rho_\s(t)$ does not have the group property in $t$. Sometimes one can show that an approximation of $\rho_\s(t)$ by a (semi-)group dynamics is possible, $\rho_\s(t)\approx e^{t\mathcal L}\rho_\s(0)$, where the generator $\mathcal L$ is called a Lindblad operator. Generically, $\mathcal L$ has eigenvalues with negative imaginary parts, which translates into irreversible dynamical behaviour appearing in the system dynamics due to the coupling with the reservoir. The question how well the true dynamics can be approximated by a semigroup dynamics | in a mathematically precise way |  is rather delicate \cite{MM20, MM22, MM23}. Heuristic derivations (without control of the error terms) can be found in many textbooks \cite{BP,RH}. The semigroup approximation is called a {\it markovian approximation} of the true dynamics. In the markovian approximation, the system dynamics $e^{t\mathcal L}\rho_\s(0)$ obeys the {\it markovian master equation} $\partial_t\rho_\s(t) = \mathcal L\rho_\s(t)$. In order for a solution of this equation to be a density matrix (non-negative operator of unit trace) for all times, the generator $\mathcal L$ must have a specific form. This is the so-called GKSL form of the generator, given by the celebrated Gorini-Kossakowski-Sudarshan-Lindblad theorem \cite{GKS,Lindblad,ChrPas}. 

We note that the dynamics of open systems generally exhibits irreversible traits (whether in the markovian approximation or not). Famous irreversible processes induced by the coupling to a reservoir are the process of decoherence, where off-diagonal density matrix elements of $\rho_\s(t)$ decay to zero as $t\rightarrow\infty$, as well as the process of thermalization, where $\rho_\s(t)$ converges to a thermal equilibrium state (Gibbs state) as $t\rightarrow\infty$ because it is in contact with a thermal reservoir \cite{TrushEtAl}.

\section{Open quantum systems by example: The open Jaynes-Cummings model}
\label{sec:jaynes}

The Jaynes-Cummings model is a paradigmatic model for an atom in an optical cavity \cite{SK}. The atom is modeled by a two-level system representing the two energetic levels of a true atom participating in exchange processes with the optical field, which is modeled by collection of harmonic oscillators. The atom is considered to be the system $\s$ and has Hilbert space $\cx^2$. We denote the ground and excited state of the atom by  $|0\rangle$ and $|1\rangle$, and we use the raising and lowering operators $\sigma_+=|1\rangle\langle0|$ and $\sigma_-=|0\rangle\langle1|$. The cavity, which takes the role of the reservoir $\r$, is described by a collection of harmonic oscillators with creation and annihilation operators $b_k^\dag$, $b_k$, $k=1,2,...$, satisfying the usual commutation relations $[b_k,b_k^\dag]=\delta_{kl}$. The Hamiltonian of the $\s\r$ complex is the sum of the system Hamiltonian, the reservoir Hamiltonian plus the interaction Hamiltonian, 
\[
H=H_\s+H_\r+H_\i,\qquad H_0= H_\s+H_\r.
\]
$H_0$, the uncoupled part, is the sum of $H_\s$ plus $H_\r$, 
\[
H_\s=\omega_0|1\rangle\langle1|=\omega_0\sigma_+\sigma_-,\qquad H_\r=\sum_k\omega_kb_k^\dag b_k,
\]
where $\omega_0>0$ is the system Bohr frequency and $\omega_k$ is the frequency of the oscillator $k$. The interaction Hamiltonian allows for exchanges of energy between $\s$ and $\r$,
\[
H_\i=\sigma_+\otimes B+\sigma_-\otimes B^\dag, \qquad B=\sum_kg_kb_k
\]
and where the complex numbers $g_k\in \cx$ constitute the `form factor'. The `size' of $g_k$ indicates how strongly the mode $k$ interacts with the atom. Let $\psi(t)=e^{-it H}\psi_0$ be the evolution of the $\s\r$ complex. We define the {\it interaction picture} dynamics by setting
\[
\phi(t) = e^{itH_0}\psi(t) = e^{itH_0}e^{-it H}\psi_0.
\]
Note that $\phi(0)=\psi_0$. The state vector $\phi(t)$ satisfies the evolution equation
\begin{equation}
\label{1}
i\partial_t \phi(t) = H_\i(t)\phi(t),\qquad H_\i(t) = e^{itH_0} H_\i e^{-itH_0}.
\end{equation}
An explicit calculation gives
\[
H_\i(t) =  \sigma_+(t)B(t)+{\rm h.c.},\qquad \mbox{with \ $\sigma_+(t) = e^{i\omega_0t}\sigma_+$ \ and\ \   $B(t)=\sum_k g_k e^{-i\omega_k t}b_k\,$}.
\]
The property which makes this model `explicitly solvable' is that the number of total excitations is conserved. Namely, the total number operator 
\[
N=|1\rangle\langle 1| +\sum_k b^\dag_kb_k
\]
commutes with the Hamiltonian,
\[
[H,N]=0.
\]
This means that the spectral subspaces of $N$ are invariant under the dynamics generated by $H$. Denote by $\vac_\r$ the state of $\r$ where all the oscillators are in their ground state and denote by $|k\rangle_\r$ the state of the oscillators in which they are all in the ground state except for the $k$th one, which is in its first excited state, that is,
\begin{align*}
|k\rangle_\r = b^\dag_k\vac_\r.
\end{align*}
We also introduce the $\s\r$ states
\begin{equation}
\label{3}
\psi_0=|0\rangle_\s\otimes \vac_\r\qquad \psi_1=|1\rangle_\s\otimes \vac_\r\qquad \chi_k=|0\rangle_\s\otimes |k\rangle_\r.
\end{equation}
States with at most one total excitation, which we will take as intial states, are of the form 
\begin{equation}
\label{2}
\phi(0)=c_0\psi_0+c_1(0)\psi_1+\sum_kd_k(0)\chi_k
\end{equation}
for some amplitudes (complex numbers) $c_0, c_1(0), d_k(0)$. By the conservation of the total number of excitations, the evolved state $\phi(t)$ in the interaction picture also has at most one excitation for all times $t$, so it is of the form
\begin{equation}\label{wavefunction}
\phi(t)=c_0\psi_0+c_1(t)\psi_1+\sum_kd_k(t)\chi_k.
\end{equation}
The dynamics in the manifold of at most one excitation is then encoded in the functions $c_1(t)$, $d_k(t)$. The atom density matrix in the interaction picture is given by 
\begin{equation}
\rho_{\s,\i}(t) :=\tr_\r\big(|\phi(t)\rangle\langle\phi(t)|\big).
\end{equation}
\begin{ex}\label{ex1}
Show that $\rho_{\s,\i}(t)$, written as a matrix in the ordered orthonormal basis $\{|1\rangle,|0\rangle\}$, has the form
\begin{equation}
\label{5}
\rho_{\s,\i}(t) :=\tr_\r\big(|\phi(t)\rangle\langle\phi(t)|\big)=\left(\begin{array}{cc}
        |c_1(t)|^2 & \overline{c_0}c_1(t) \\[4pt]
        c_0\overline{c_1(t)} & 1-|c_1(t)|^2
    \end{array}\right),
\end{equation}
where the bar denotes the complex conjugate.
\end{ex}

We now derive the differential equation satisfied by $\rho_{\s,\i}$. The fraction$\frac{\partial_t c_1(t)}{c_1(t)}$ is a complex number with real and imaginary parts written as 
\begin{equation}
\label{4}
\frac{\partial_tc_1(t)}{c_1(t)} = -\tfrac12\gamma(t)-i\tfrac12 S(t)
\end{equation}
so that 
\[
c_1(t) =e^{-\frac12 \int_0^t (\gamma(\tau) -i S(
\tau) )d\tau} c_1(0).
\]
We assume that $c_1(0)\neq 0$. We have $\partial_t|c_1(t)|^2=-\gamma(t)|c_1(t)|^2$ and taking the time derivative of the above equation \eqref{5} for $\rho_{\s,\i}(t)$ and using \eqref{4} gives
\begin{equation}
\label{ny1}
\partial_t \rho_{\s,\i}(t) = 
\begin{pmatrix}
-\gamma(t) |c_1(t)|^2 & -\tfrac12 \overline{c_0}\big(\gamma(t) +i S(t)\big)c_1(t)\\[4pt]
-\tfrac12 c_0\big(\gamma(t)-iS(t)\big)\overline{c_1(t)} & \gamma(t) |c_1(t)|^2
\end{pmatrix}.
\end{equation}
\begin{ex}\label{ex2}
Show that \eqref{ny1} can be written as
\begin{equation}
\label{ny2}
\partial_t\rho_{\s,\i}(t) = -\tfrac i2 S(t)\big[\sigma_+\sigma_-\, , \, \rho_{\s,\i}(t) \big] +\gamma(t)\Big( \sigma_-\rho_{\s,\i}(t) \sigma_+ -\tfrac12 \big\{ \sigma_+\sigma_-\, ,\, \rho_{\s,\i}(t)\big\}\Big),
\end{equation}
where $[x,y]=xy-yx$ is the commutator and $\{x,y\}=xy+yx$ is the anti-commutator of two operators $x,y$. 
\end{ex}
By undoing the interaction picture, $\rho_\s(t) =e^{-itH_\s}\rho_{\s,\i}(t)e^{itH_\s}$, we obtain from \eqref{ny2} 
\[
\partial_t\rho_\s(t) = -i\big[ H_\s+\tfrac 12 S(t)\sigma_+\sigma_-\, , \, \rho_\s(t) \big] -\gamma(t)\Big( \sigma_-\rho_\s(t) \sigma_+ -\tfrac12 \big\{ \sigma_+\sigma_-\, ,\, \rho_\s(t)\big\}\Big).
\]
Let us investigate $c_1(t)$ in more detail. We project the equation \eqref{1} on $\langle \psi_1|$ and $\langle \chi_k|$ (recall \eqref{3}) to obtain
\begin{align*}
i\partial_tc_1(t) &= \langle \psi_1, H_\i(t)\phi(t)\rangle =\langle\psi_1, \sigma_+(t)B(t)\Phi(t)\rangle =\sum_k g_k e^{i(\omega_0-\omega_k)t} d_k(t),\\
i\partial_t d_k(t) &= \overline{g_k}\, e^{-i(\omega_0-\omega_k)t}c_1(t).
\end{align*}
Assume that $d_k(0)=0$. Then $d_k(t) = -i\, \overline{g_k} \int_0^t e^{-i(\omega_0-\omega_k)s}c_1(s) ds$ and using this in the equation above for $\partial_t c_1(t)$ gives
\begin{equation}
\label{ny3}
\partial_t c_1(t) = -\int_0^t f(t-s) c_1(s) ds,
\end{equation}
where 
\begin{equation}
\label{ny5}
f(\tau) = \sum_k|g_k|^2 e^{-i\omega_k\tau} e^{i\omega_0 \tau} = \tr_\r\big(\vac\vacbra\, B(\tau) B^\dag\big) e^{i\omega_0 \tau}.
\end{equation}
The quantity $\tr_\r\big(\vac\vacbra\, B(\tau) B^\dag\big)$ is called the {\it vacuum correlation function} (here, of course, $B^\dag=B(0)^\dag$). 
\medskip

We now perform the {\it continuous mode limit.} We write
\[
\sum_k |g_k|^2 e^{-i\omega_k\tau} = \sum_\omega e^{-i\omega\tau} \sum_{\{k\,:\,\omega_k=\omega\}} |g_k|^2 = \sum_\omega e^{-i\omega\tau} n(\omega)|g(\omega)|^2
\]
where we consider $g_k=g(\omega_k)$ to be a function of $\omega_k$ and $n(\omega)$ is the number of oscillators having the frequency $\omega$. Denoting by $\Delta\omega$ the gap between consecutive frequencies of the oscillators we get
\[
\sum_k |g_k|^2 e^{-i\omega_k\tau} =\sum_\omega \frac{n(\omega)}{\Delta\omega} e^{-i\omega\tau} |g(\omega)|^2 \Delta\omega\longrightarrow \int_0^\infty \varrho(\omega) |g(\omega)|^2 e^{-i\omega\tau}d\omega \equiv C(\tau)
\]
where $\varrho(\omega)$ is the density of modes and $C(\tau)$ is the reservoir correlation function. In the first equality above, we rearranged the sum: instead of summing over all $k$ we sum over all $\omega$, so we need to introduce the multiplicity $n(\omega)$; we then divide and multiply by $\Delta\omega$ to make the density of modes and the integration measure apparent.

\begin{figure}
  \centering
   \includegraphics[width=0.5\linewidth]{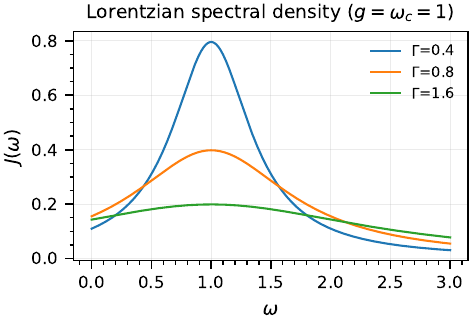}
   \caption{Lorentzian spectral density  \eqref{ny4} for different values of $\Gamma$.}
   \label{fig:Fig-Lorentzian}
\end{figure}

By the Riemann-Lebesgue Lemma, $C(\tau)$ typically decays to zero as $\tau\rightarrow\infty$, at a speed depending on the regularity (smoothness) of $\varrho(\omega)|g(\omega)|^2$. The correlation function is the Fourier transform of the {\it spectral density} defined as
\[
J(\omega) = \sqrt{2\pi}\varrho(\omega)|g(\omega)|^2,\qquad \omega\ge0.
\]
For concrete models (geometries of the cavity) one can derive the form of the spectral density. A popular form is the Lorentzian  centered at $\omega_c>0$ with width $\Gamma>0$,
\begin{equation}
\label{ny4}
J(\omega) = \frac{g}{\pi}\frac{\Gamma}{(\omega-\omega_c)^2 +\Gamma^2},
\end{equation}
where $g>0$; see \cite{Garraway} and also Figure \ref{fig:Fig-Lorentzian}.
Combining \eqref{ny5} with \eqref{ny4} gives, 
\begin{equation}
\label{ny7}
f(\tau)=\dfrac{1}{\sqrt{2\pi}}\int_{-\infty}^{\infty}J(\omega)e^{-i(\omega-\omega_0)\tau}d\omega.
\end{equation}

\begin{ex}
\label{exercise3}
Show that $f(\tau)$ given in \eqref{ny7} with the spectral density $J(\omega)$ given in \eqref{ny4}, is 
\[
f(\tau)=\frac{g}{\sqrt{2\pi}}e^{-(\Gamma-i\Delta)\tau},
\]
where the {\it detuning} 
\[
\Delta=\omega_0-\omega_c
\]
is the difference between the atomic and cavity central frequencies. Then solve the differential equation \eqref{ny3} to get the solution
\begin{equation}\label{solutionDE}
c_1(t) = c_1(0) e^{-(\Gamma-i\Delta)t/2} \Big[\cosh(Rt/2) +\frac{\Gamma-i\Delta}{R}\sinh(Rt/2)\Big],
\end{equation}
where
\begin{equation}
\label{Rdef}
R=\sqrt{(\Gamma-i\Delta)^2-\frac{4g}{\sqrt{2\pi}}} \qquad \text{(principal branch)}
\end{equation}
(see also \cite{LiZouShao}).
\end{ex}

In the {\it resonant case} $\Delta=0$ we get
\[
c_1(t) = c_1(0) e^{-\Gamma t/2}\Big[\cosh(R_0t/2) +\frac{\Gamma}{R_0}\sinh(R_0t/2)\Big],\qquad R_0=\sqrt{\Gamma^2 -\frac{4g}{\sqrt{2\pi}}}\, .
\]
For $\Gamma>4g/\sqrt{2\pi}$ we have $R_0>0$ and so $c_1(t)$ is real, meaning that $S(t)=0$ in \eqref{4} and 
\[
c_1(t) = c_1(0) e^{-\frac12\int_0^t\gamma(\tau)d\tau},\qquad \gamma(t) = \frac{4g}{\sqrt{2\pi}}\, \frac{\sinh(R_0t/2)}{R_0\cosh(R_0t/2)+\Gamma \sinh(R_0t/2)}\, .
\]
In the short and the long time regimes we find the following behavior:
\begin{align*}
R_0t/2 <\!\!<1: & \quad \gamma(t)\approx \frac{4g}{\sqrt{2\pi}} \frac{t}{2+\Gamma t}\approx \frac{2g}{\sqrt{2\pi}} \,t \ \quad \text{($\Gamma t/2<\!\!<1$ for the last approximation)}\\
R_0t/2 >\!\!>1: & \quad \gamma(t)\approx \frac{4g}{\sqrt{2\pi}} \frac{1}{R_0+\Gamma}\approx \frac{2g}{\sqrt{2\pi}} \frac1\Gamma \ \quad \text{($g<\!\!<\Gamma$ for the last approximation)}
\end{align*}
This shows that $c_1(t)$ decays (approximately) exponentially in time for large times $c_1(t)\approx c_1(0) e^{-gt/(\sqrt{2\pi}\Gamma)}$ while for small times it decays as $c_1(t)\approx c_1(0) e^{-gt^2/(2\sqrt{2\pi}\Gamma)}$. 
\medskip

{\bf Summary.} The reduced atom dynamics shows irreversibility due to the coupling with a continuum of oscillator modes. Namely, as $t\rightarrow\infty$ we have $\rho_\s(t)\rightarrow |0\rangle\langle0|$, the atom relaxes to its ground state. The initial excitation in the atom is `radiated off' into the reservoir. The speed of convergence of this process is initially as $e^{-\alpha t^2}$ and for large times it is exponential, $e^{-\alpha't}$. We have derived the exact evolution equation for the atom,
\begin{equation}
\label{exevol}
\partial_t\rho_\s(t) = -i\big[H_{\rm eff}(t)\, ,\, \rho_\s(t)\big] +\gamma(t) \big(\sigma_-\rho_\s(t)\sigma_+ -\tfrac12\big\{\sigma_+\sigma_-\, ,\, \rho_\s(t)\big\}\big).
\end{equation}
In the resonant case the effective Hamiltonian is simply $H_{\rm eff}(t) = H_\s=\omega_0\sigma_+\sigma_-$. For large times we have $\gamma(t)\approx\gamma$ constant and the evolution equation becomes,
\[
\partial_t\rho_\s(t) = -i\big[H_\s\, ,\, \rho_\s(t)\big] +\gamma \mathcal D[\rho_\s(t)]= \ml [\rho_\s(t)],
\]
where $\mathcal D$ is called the {\it dissipator} and $\ml$ is the {\it Lindblad operator}. They are both superoperators, that is, operators acting on density matrices. The last equation is of markovian (semigroup) form and the solution can be written as  $\rho_\s(t)=e^{t\ml}\rho_\s(0)$. 
\medskip

The above discussion leads to the following question:
Suppose $\mathcal L$ is a superoperator of a quantum system. What are the properties of $\ml$ such that for any $\rho_0$ (`initial state'), $e^{t\ml}\rho_0$ is a density matrix for all times $t\ge 0$? Such $\ml$ are the generators of quantum dynamical semigroups. We are going to identify the form of such generators in Section \ref{sec:QDSG}. Before addressing that issue, we are analyzing an important class of maps, called CPTP maps.

\section{CPTP maps}
\label{sec:cptpmaps}

Quite generally, one may describe a `process' of (an `effect' on) a quantum system having Hilbert space $\h$, as a transformation of its state, where an incoming state is mapped to an outgoing state, according to a mapping $\Phi: {\mathcal B}(\h)\rightarrow\mathcal B(\h)$. This mapping should yield, for every density matrix $\rho\in\mathcal B(\h)$, another density matrix $\Phi(\rho)$. In particular, it should map positive operators to positive operators and it should preserve the trace. Such a map is called positive and trace preserving (PTP). Now if we think of a system which is composed of several subsystems (say a chain of spins or a register of several qubits), one may look at processes which only influence some of the subsystems, not others. Say you have two qubits, $Q_1$ and $Q_2$ and you affect (by some measurement, interaction) only qubit $Q_1$, not $Q_2$. Denote the (PTP) map corresponding to the action on qubit $Q_1$ by $\Phi_1:\mathcal B(\cx^2)\rightarrow\mathcal B(\cx^2)$. Then $\Phi = \Phi_1\otimes\bbbone_{\mathcal B(\cx^2)}$ is the PTP map acting on both qubits, with only $Q_1$ affected according to $\Phi_1$. Does $\Phi$ map density matrices of two qubits to density matrices of two qubits? The answer is, generally {\it no}! Of course, if $\rho_{12}=\rho_1\otimes\rho_2$ is not entangled, then yes, $\Phi(\rho_1\otimes\rho_2)=\Phi_1(\rho_1)\otimes\rho_2$ is still positive. But in general, if $\rho_{12}$ is entangled, then $\Phi(\rho_{12})$ might be a matrix which is not positive semidefinite! A famous example is this: take for $\Phi_1$ the PTP map which takes the transpose of a matrix. Then $\Phi=\Phi_1\otimes\bbbone_{\mathcal B(\cx^2)}$ is called the partial transpose (map). By what we just discussed, if $\rho_{12}=\rho_1\otimes\rho_2$ is not entangled, then $\Phi(\rho_{12})$ is positive semidefinite. So if $\Phi(\rho_{12})$ has some negative eigenvalues, then you know that the density matrix $\rho_{12}$ must be entangled. This simple fact is the core of the so-called Peres-Horodecki, or PPT criterion (positive partial transpose): A sufficient condition of entanglement in a bipartite system is that the partial transpose of the state have a negative eigenvalue. For qubits, the condition is also necessary \cite{HHH}, and it has applications even in infinite dimensional settings \cite{MZ23}. 

We have digressed a bit, but we have found out that it is not sufficient for $\Phi_1$ to be positivity preserving in order to guarantee that $\Phi=\Phi_1\otimes\bbbone_{\mathcal B(\cx^2)}$ be positivity preserving | an effect due to the possibility of entanglement. If we want $\Phi$ to map density matrices to density matrices, we must thus require more on $\Phi_1$ than simply preservation of positivity. The stronger condition is called {\it complete positivity} of $\Phi_1$, meaning that $\Phi_1\otimes \bbbone_{\mathcal B(\cx^N)}$ be positivity preserving for any $N$. (The dimension $N$ here is arbitrary, as the above argument should hold when $Q_1$ is coupled to any other system of dimension $N$.) 

These considerations motivate the  study of CPTP maps. In this section, we will show that the interaction with a reservoir results in a CPTP map on the system. We then demonstrate in Theorem 1 below that any CPTP map has a special structure, called the Kraus representation. Using the Kraus representation, we are then able to show in Theorem 2 that a map on a system is CPTP if and only if it derives from an interaction with a reservoir. This provides a beautiful link between the theory of CPTP maps and the theory of open quantum systems.

\bigskip

Let $\h$ and $\h_\r$ be two Hilbert spaces (of pure states) and let $\Omega\in\h_\r$ be a pure state (normalized vector). Let $U$ be a unitary on the tensor product $\h\otimes\h_\r$. Then define the map $\Phi$ acting on $\mathcal B(
\h)$, the bounded operators on $\h$, by
\begin{equation}
\label{Phi}
\Phi(X) = \tr_\r\big[U (X\otimes |\Omega\rangle\langle\Omega|)U^\dag\big],\qquad X\in\mathcal B(\h).
\end{equation}
Here, $\tr_\r$ is the partial trace over the Hilbert space $\h_\r$. It is defined as follows. For an operator $X=A\otimes B$ acting on $\h\otimes\h_\r$ we have 
\[
\tr_\r (A\otimes B) = A\, \tr(B).
\]
The right hand side is the operator $A$ multiplied by the number $\tr(B)$ (the `ordinary' trace of $B$). We then extend the action of $\tr_\r$ by linearity to all of $\mathcal B(\h\otimes\h_\r)$. Note that $\tr_\r X$ is still an operator, acting on $\h$. In practice, to find the partial trace of a given $X$ you first write $X$ as a linear combination of product operators, $X=\sum_i A_i\otimes B_i$. Then $\tr_\r X = \sum_i A_i\, \tr(B_i)$. 

The unitary $U$ can be written as $U=\sum_{ij} u_{ij} P_i\otimes Q_i$ with $u_{ij}\in\cx$ and $P_i,Q_j$ operators on $\h, \h_\r$ (here we assume the dimensions of $\h,\h_\r$ to be finite, so as not to have to worry about convergence of the sums). Then
\begin{align*}
\Phi(X) & =\tr_\r\big[ U\big(X\otimes |\Omega\rangle\langle\Omega|\big) U^\dag\big]\\
&=\sum_{ijkl}u_{ij}\overline{u_{kl}}\, \tr_\r \big[(P_i\otimes Q_j)(X\otimes |\Omega\rangle\langle \Omega|) (P^\dag_k\otimes Q^\dag_l)\big]\\
&=\sum_{ijkl} u_{ij}\overline{u_{kl}}\, P_iXP^\dag_k\, \tr\big[Q_j|\Omega\rangle\langle\Omega|Q^\dag_l\big].
\end{align*}
Let $\{f_\alpha\}$ be an orthonormal basis of $\h_\r$. The trace of an operator $B$ on $\h_\r$ is given by $\tr(B)=\sum_\alpha \langle f_\alpha, Bf_\alpha\rangle$. It follows that 
\begin{align*}
\Phi(X) & =\sum_\alpha\Big(\sum_{ij} u_{ij}\langle f_\alpha, Q_j\Omega\rangle P_i\Big) X \Big(\sum_{kl} \overline{u_{kl}}\langle \Omega, Q^\dag_l f_\alpha\rangle P_k^\dag\Big).
\end{align*}
The operator to the right of $X$ is the adjoint of the operator to the left and so we can write
\begin{equation}
\label{PhiK}
\Phi(X) = \sum_\alpha K_\alpha X K^\dag_\alpha,\qquad K_\alpha\equiv \sum_{ij}u_{ij} \langle f_\alpha, Q_j\Omega\rangle P_i.
\end{equation}
The operators $K_\alpha$ are called  {\it Kraus operators}. They act on $\h$. Let us investigate some properties of the Kraus operators. Firstly,
\[
\tr\,\Phi(X) = \tr_{\h\otimes\h_\r} \big[ U (X\otimes |\Omega\rangle\langle\Omega|)U^\dag\big] =\tr_{\h\otimes\h_\r} \big[X\otimes |\Omega\rangle\langle\Omega|\big]=\tr X  
\]
by the cyclicity of the trace and the unitarity of $U$. So $\Phi$ is trace preserving. Secondly, 
$\tr\Phi(X) = \tr\big(\sum_\alpha K_\alpha XK^\dag_\alpha\big) = \tr\big(\sum_\alpha K^\dag_\alpha K_\alpha X\big)$ which together with the trace preserving property gives, $\tr\big(\sum_\alpha K^\dag_\alpha K_\alpha -\bbbone\big) X=0$ for all $X\in\mathcal B(\h)$. This implies that 
\[
\sum_\alpha K^\dag_\alpha K_\alpha=\bbbone, 
\]
as is shown in the following exercise.

\begin{ex}\label{exercise4}
    Let $\mathcal{H}$ be a finite-dimensional Hilbert. Show that if $A\in\mathcal{B}(\mathcal{H})$ (bounded operators on $\mathcal{H}$) is such that $\tr(AX)=0$ $\forall X\in\mathcal{B}(\mathcal{H})$, then $A=0$.
\end{ex}

Next recall that an operator $A$ is called positive, denoted $A\ge 0$ if $A^\dag=A$ and the spectrum | all eigenvalues $a$ of $A$ | satisfy $a\ge 0$. Equivalently, $A\ge 0$ if and only if $A=B^\dag B$ for some operator $B$. Equivalently, $A\ge 0$ if and only if $\langle x,Ax\rangle\ge 0$ for all $x\in\h$. 

For an integer $n\ge 1$ consider the map $\Phi\otimes\bbbone_{n\times n}$ acting on $\h\otimes\mathcal B(\cx^n)$. If $A\in\mathcal B(\h\otimes \cx^n)$ is a positive operator, $A\ge 0$, then $(\Phi\otimes \bbbone_{n\times n}) (A) =\sum_\alpha (K_\alpha\otimes\bbbone_{n\times n}) A (K^\dag_\alpha\otimes \bbbone_{n\times n})$ is also a positive operator.  This means that $\Phi\otimes\bbbone_{n\times n}$ is a positivity preserving map for all $n\ge 1$. A map $\Phi$ such that $\Phi\otimes \bbbone_{n\times n}$ is positivity preserving for all $n\ge 1$ is called a {\it completely positive (CP) map}. What we have shown is that  $\Phi$ given in \eqref{Phi} or \eqref{PhiK}, is a completely positive and trace preserving, CPTP, map.

We are now going to show the amazing fact that {\it every} CPTP map is of the Kraus form $\sum_\alpha K_\alpha (\cdot) K^\dag_\alpha$ with operators satisfying $\sum_\alpha K^\dag_\alpha K_\alpha=\bbbone$. Let $\Phi$ be CPTP on $\mathcal B(\h)$. Take a $\psi\in\h$ and investigate $\Phi(|\psi\rangle\langle\psi|)$. Pick an orthonormal basis $\{e_i\}$ of $\h$ and let $\c$ be the anti-linear map taking complex conjugation of components of vectors in this basis. Then $\c^2=\bbbone$, $\c e_i=e_i$ and for any $\phi,\psi\in\h$, $\langle \c\psi,\c\phi\rangle=\langle\phi,\psi\rangle$. Expanding $\psi$ in the basis as $\psi=\sum_j\langle e_j,\psi\rangle e_j$ and using $\langle e_j,\psi\rangle=\langle \c \psi,e_j\rangle$ and $\overline{\langle e_k,\psi\rangle} = \langle e_k,\c\psi\rangle$ we get
\begin{align*}
\Phi\big(|\psi\rangle\langle\psi|\big) &= \sum_{jk}\langle e_j,\psi\rangle \Phi\big(|e_j\rangle\langle e_k|\big) \overline{\langle e_k,\psi\rangle}
= \sum_{jk}\langle \c \psi,e_j\rangle\langle e_k,\c\psi\rangle \Phi\big(|e_j\rangle\langle e_k|\big) \\
&=\sum_{jk}\tr_2\Big[\big(\Phi\otimes |\c\psi\rangle\langle\c\psi|\big) (|e_j\rangle\langle e_k|\otimes |e_j\rangle\langle e_k|)\Big]
\end{align*}
where in the last equality we have doubled the Hilbert space, $\h\otimes\h$ and $\tr_2$ is the partial trace over the second factor. Now $|e_j\rangle\langle e_k|\otimes |e_j\rangle\langle e_k|=|e_j\otimes e_j\rangle\langle e_k\otimes e_k|$ and setting $v=\sum_j e_j\otimes e_j$ we obtain,
\[
\Phi\big(|\psi\rangle\langle \psi|\big) = \tr_2\Big[\big(\Phi\otimes |\c\psi\rangle\langle\c \psi|\big)(|v\rangle\langle v|)\Big].
\]
Next, since $\Phi$ is CP, the operator $(\Phi\otimes\bbbone)(|v\rangle\langle v|)$ is positive and so $(\Phi\otimes\bbbone)(|v\rangle\langle v|) = \sum_\alpha |s_\alpha\rangle\langle s_\alpha|$ for some $s_\alpha\in\h\otimes\h$ (this follows for example from the spectral theorem). Therefore,
\begin{equation}
\label{x}
\Phi\big(|\psi\rangle\langle \psi|\big) = \sum_\alpha \tr_2\Big[\big(\bbbone\otimes |\c\psi\rangle\langle\c \psi|\big)|s_\alpha\rangle\langle s_\alpha|\Big].
\end{equation}
Fix $\alpha$ and write $|s\rangle$ for $|s_\alpha\rangle$ for the moment. We have $s=\sum_{mn} z_{mn} e_m\otimes e_n$ for some coordinates $z_{mn}\in\cx$ and 
\begin{align*}
\tr_2 \Big[\big(\bbbone\otimes |\c\psi\rangle\langle\c \psi|\big)|s\rangle\langle s|\Big] &= 
\sum_{mnpq}z_{mn}\overline{z_{pq}} \, \tr_2\Big[ \big(\bbbone\otimes |\c\psi\rangle\langle\c \psi|\big)|e_m\otimes e_n\rangle\langle e_p\otimes e_q|\Big]\\
&= \sum_{mnpq}z_{mn}\overline{z_{pq}}\, \langle \c\psi,e_n\rangle\langle e_q,\c\psi\rangle \, |e_m\rangle\langle e_p|\\
&= \Big(\sum_{mn}z_{mn}|e_m\rangle\langle e_n,\psi\rangle \Big) \Big(\langle\psi| \sum_{pq}\overline{z_{pq}} |e_q\rangle\langle e_p|\Big)
\end{align*}
where we have used that $|e_m\otimes e_n\rangle\langle e_p\otimes e_q| = |e_m\rangle\langle e_p|\otimes |e_n\rangle\langle e_q|$ in the second step and $\langle \c\psi,e_n\rangle=\langle e_n,\psi\rangle$, $\langle e_q,\c\psi\rangle = \langle \psi,e_q\rangle$ in the third step. Introducing the operator $K=\sum_{mn}z_{mn}|e_m\rangle\langle e_n|$ the above shows that $\tr_2 \big[\big(\bbbone\otimes |\c\psi\rangle\langle\c \psi|\big)|s\rangle\langle s|\big]=|K\psi\rangle\langle K\psi|$. Remember that the $s$ stands for a general but fixed $s_\alpha$ | together with \eqref{x} we have shown that there are operators $K_\alpha$ such that for all $\psi\in\h$,
\[
\Phi\big(|\psi\rangle\langle \psi|\big) = \sum_\alpha K_\alpha |\psi\rangle\langle\psi| K^\dag_\alpha. 
\]
The action of $\Phi$ on rank-one projection operators $|\psi\rangle\langle\psi|$ has the above Kraus form. How about $\Phi(X)$ for a general $X\in\mathcal B(\h)$? We decompose $X={\rm Re}X +i{\rm Im}X$ where ${\rm Re}X=\tfrac12(X+X^\dag)$ and ${\rm Im} X = \tfrac{1}{2i}(X-X^\dag)$ are hermitian operators. As $\Phi$ is a linear map, $\Phi(X)=\Phi({\rm Re}X)+ i\Phi({\rm Im}X)$. In its diagonal form we have ${\rm Re}X = \sum_j \lambda_j |\psi_j\rangle\langle \psi_j|$ with eigenvalues $\lambda_j\in\rx$ and eigenvectors $\psi_j$. By the linearity of $\Phi$ and by the known action in Kraus form on projections $|\psi_j\rangle\langle\psi_j|$, we have $\Phi({\rm Re}X) = \sum_j\lambda_j \sum_\alpha K_\alpha|\psi_j\rangle\langle \psi_j|K^\dag_\alpha = \sum_\alpha K_\alpha ({\rm Re}X)K^\dag_\alpha$. In the same way $\Phi({\rm Im}X) = \sum_\alpha K_\alpha ({\rm Im}X)K^\dag_\alpha$. We conclude that  $\Phi(X) = \sum_\alpha K_\alpha XK^\dag_\alpha$, $\forall X\in\mathcal B(\h)$. Notice that since we started off with a CPTP $\Phi$ we have automatically $\sum_\alpha K_\alpha^\dag K_\alpha=\bbbone$ (by the same argument as above). We have shown the following result.

\begin{thm}[Kraus representation theorem]
Suppose $\Phi$ is CPTP on $\mathcal B(\h)$, $\dim\h=d<\infty$. Then there are operators $K_\alpha\in\mathcal B(\h)$, $\alpha=1,\ldots,d^2$ such that $\sum_\alpha K^\dag_\alpha K_\alpha=\bbbone$ and for all $X\in\mathcal B(\h)$,
\[
\Phi(X) = \sum_\alpha K_\alpha X K^\dag_\alpha.
\]
Conversely, if $\Phi$ is a map on $\mathcal B(\h)$ having this form for some operators $K_\alpha$ with $\sum_\alpha K_\alpha^\dag K_\alpha=\bbbone$, then $\Phi$ is CPTP.
\end{thm}
While we have stated and proven this theorem in the finite dimensional case, there are ways to extend it to infinite-dimensional Hilbert space \cite{Davies,Attal}. 
\medskip

The equivalence for a $\Phi$ to be CPTP and of Kraus form is a beautiful result | and it is even more beautiful because there is a link to open quantum system theory as we explain now. Recall the original definition of $\Phi$,
\[
\Phi(X) = \tr_\r\Big[ U \big(X\otimes |\Omega\rangle\langle \Omega|\big)U^\dag\Big].
\]
The right side has a physical interpretation, which we shall call the `open system representation' of $\Phi$. $U$ can be viewed as the dynamics of a bipartite system with Hilbert space $\h\otimes\h_\r$ | for instance, $U=e^{-itH}$ for a fixed time $t$ where $H$ is a system-reservoir Hamiltonian. The reservoir is in the state $\Omega\in\h_\r$ and $X=\rho$ can represent a system density matrix. Then $\Phi(X)$ is the reduced density matrix of the system after it has interacted (by $U$) with the reservoir. We have shown so far that this special structure, having an initial product bipartite state ($X\otimes |\Omega\rangle\langle \Omega|$), then acting with an overall (generally coupling) dynamics $U$, then reducing the state to the system alone, leads to a CPTP map. Now we show that the converse is true as well! Namely, given any CPTP map $\Phi$ on $\mathcal B(\h)$ we can construct a reservoir with Hilbert space $\h_\r$ and state $\Omega\in\h_\r$, and a unitary $U$ on $\h\otimes\h_\r$, such that $\Phi$ has the associated physical representation as the reduction of the full state to the system alone. 

\begin{thm}[Open systems representation of CPTP maps | dilation theorem]
\label{thm2}
Let $\Phi$ be a linear map on $\mathcal B(\h)$ for a Hilbert space $\h$ with $\dim\h=d<\infty$. The following are equivalent:
\begin{itemize}
\item[\rm (1)] $\Phi$ is CPTP.
\item[\rm (2)] There exists a Hilbert space $\h_\r$ of dimension $\le d^2$, a unit vector $\Omega\in\h_\r$ and a unitary $U$ on $\h\otimes\h_\r$, such that 
\[
\Phi(X) = \tr_\r\Big[ U \big(X\otimes |\Omega\rangle\langle \Omega|\big)U^\dag\Big]\qquad \forall X\in\mathcal B(\h).
\]
\end{itemize}
\end{thm}

\noindent
We have already shown above that $(2)\Rightarrow (1)$. To show the converse we take a CPTP map $\Phi$ on $\mathcal B(\h)$ and we construct a Hilbert space $\h_\r$, a unit vector $\Omega\in\h_\r$ and a unitary $U$ on $\h\otimes\h_\r$ such that $\Phi$ has the desired form as in (2). As $\Phi$ is CPTP it has a Kraus representation $\Phi(X)=\sum_{\alpha=1}^N K_\alpha XK_\alpha^\dag$ where $N\le d^2$ with $d=\dim\h$. We take $\h_\r=\cx^N$ and let $\{e_\alpha\}_{\alpha=1}^N$ be an orthonormal basis of $\h_\r$. For $\psi,\chi\in\h$,
\begin{align*}
\Phi(\big(|\psi\rangle\langle\chi|\big) &= \sum_\alpha K_\alpha |\psi\rangle\langle\chi| K^\dag_\alpha =\sum_\alpha \tr_\r \Big[K_\alpha|\psi\rangle\langle\chi| K^\dag_\alpha\otimes |e_\alpha\rangle\langle e_\alpha|\Big]\\
&=\sum_\alpha \tr_\r |K_\alpha\psi\otimes e_\alpha\rangle\langle K_\alpha\chi\otimes e_\alpha|.
\end{align*}
We would like this to be of the form $\tr_\r \big[U|\psi\otimes\Omega\rangle\langle\chi\otimes\Omega|U^\dag\big]$. Pick any $\Omega\in\h_\r$, $\|\Omega\|=1$ and define the linear map $\bar U:\h\otimes\cx\Omega\rightarrow\h\otimes\h_\r$ by
\[
\bar U\psi\otimes\Omega = \sum_\alpha K_\alpha\psi\otimes e_\alpha.
\]
Then we have
\begin{align*}
\tr_\r \big[\bar U \big( |\psi\rangle\langle \chi|\otimes|\Omega\rangle\langle \Omega|\big) \bar U^\dag \big] &= \tr_\r |\bar U\psi\otimes\Omega\rangle\langle\bar U\chi\otimes\Omega| =\sum_{\alpha,\beta} \tr_\r |K_\alpha\psi\otimes e_\alpha\rangle \langle K_\beta \chi\otimes e_\beta|\\
&=\sum_{\alpha,\beta} |K_\alpha\psi\rangle\langle K_\beta\chi| \ \tr_\r|e_\alpha\rangle\langle e_\beta|=\sum_\alpha K_\alpha|\psi\rangle\langle\chi|K^\dag_\alpha = \Phi\big(|\psi\rangle\langle\chi|\big).
\end{align*}
By linearity we obtain
\[
\tr_\r \big[ \bar U\big( X\otimes |\Omega\rangle\langle\Omega|\big) \bar U^\dag \big]= \Phi(X),\qquad \forall  X\in\mathcal B(\h).
\]
Next, 
\begin{align*}
\langle \bar U\psi\otimes\Omega, \bar U \chi\otimes\Omega\rangle &= \sum_{\alpha,\beta} \langle K_\alpha\psi\otimes e_\alpha, K_\beta \chi\otimes e_\beta\rangle = \sum_\alpha \langle\psi, K^\dag_\alpha K_\alpha\chi\rangle = \langle \psi,\chi\rangle = \langle \psi\otimes\Omega, \chi\otimes \Omega\rangle,
\end{align*}
which shows that $\bar U:\h\otimes\cx\Omega\rightarrow\h\otimes\h_\r$ preserves the inner product. 

\begin{ex}\label{exercise5}
Suppose $M$ is a subspace of a Hilbert space $\mathcal{H}$ with $\dim\mathcal{H}=N<\infty$, and that $\mathcal{U}:M\to\mathcal{H}$ is a linear map such that $\langle\mathcal U x,\mathcal U y\rangle=\langle x,y\rangle$, $\forall x,y\in M$. Show there exists a unitary $\overline{\mathcal U}$ on $\mathcal H$ such that $\overline{\mathcal U}|_M=\mathcal U$ (here $|_M$ denotes the restriction).
\end{ex}
It follows from this exercise (with $M=\h\otimes\cx\Omega$) that $\bar U$ can be extended to a unitary $U$ on $\mathcal H$. This shows that $(1)\Rightarrow (2)$ and the proof of Theorem \ref{thm2} is complete. 
\smallskip

The `dilation' of a CPTP map | namely its representation as a reduction of a map acting on a larger space | is true in much more generality. The Stinespring dilation theorem \cite{Stinespring} says the following. Let $\Phi$ be a linear map from a unital $C^*$-algebra $\mathcal A$ to $\mathcal B(\h)$, the bounded operators on a Hilbert space $\h$. Then $\Phi$ is completely positive if and only if there exists a Hilbert space $\mathcal K$, a bounded linear transformation $V:\h\rightarrow\mathcal K$ and a $*$representation $\pi$ from $\mathcal A$ into operators $\mathcal B(\mathcal K)$ such that $\Phi(A) = V^\dag\pi(A)V$ for all $A\in\mathcal A$.

\section{Quantum dynamical semigroups}
\label{sec:QDSG}

As we have seen at the end of Section \ref{sec:jaynes}, on the example of the dissipative Jaynes-Cummings model, the dynamics of an open system follows an evolution equation which contains a Hamiltonian part and a dissipative part (see equation \eqref{exevol} and discussion thereafter). Generally the `renormalized Hamiltonian' ($H_{\rm eff}(t)$) becomes time dependent and the dissipative part reflects the fact that the system can be driven to a final state, encoding irreversible effects the reservoir imposes on the system. This dynamical behaviour is different from a  closed system dynamics of a system which is not coupled to a reservoir. There the dynamics is simply given by a unitary group $e^{-itH} \cdot e^{-it H}$ (acting on density matrices), with a time independent Hamiltonian $H$ and without any dissipative contribution, resulting in an eternally oscillating behaviour in time (no relaxation). Then the question is: can we approximate the dynamics of an open system by a mapping which has the group property in time $t$, that is, is there a (super-)operator $\mathcal L$ such that $\rho_\s(t)\approx e^{t\mathcal L}\rho_\s(0)$?  As alluded to in Section \ref{sec:formalism}, this is physically reasonable to expect the validity of the Markovian approximation in a regime where the interaction is weak and the reservoir loses its memory quickly. Proving such an approximation and controlling the deviation of the approximate dynamics from the true one, is a hard task \cite{MM20, MM22, MM23}. We do not address this question here. Rather, we are asking: If such an operator $\mathcal L$ exists, what are its properties? As it turns out, $\mathcal L$ must have a specific form, called the GKSL structure, as given in Theorem \ref{thm3}, which is the main result of this section.

\medskip

Let $\h=\h_\s\otimes\h_\r$ with $\dim \h_\s=d<\infty$ and let $H=H^\dag\in\mathcal B(\h)$ and a density matrix $\rho_\r$ on $\h_\r$ be given. For $t\in\rx$ the map $\Phi_t$ defined by 
\[
X\mapsto \Phi_t(X) = \tr_\r\big(e^{-itH}(X\otimes\rho_\r)e^{itH}\big)
\]
is a CPTP map on $\mathcal B(\h)$ and $\Phi_{t=0}={\rm id}$ is the identity. By definition, a family $\{\Phi_t\}_{t\ge 0}$ of CPTP maps satisfying $\Phi_0={\rm id}$  is called a {\it CPTP dynamical map}. Since the two parts (system and reservoir) generally interact via $H$, we have $\Phi_{t+s}\neq \Phi_t\circ\Phi_s$. However, when the group property is satisfied, then we have the following definition. 
\begin{itemize}
\item[] 
A CPTP family $\{\Phi_t\}_{t\ge 0}$ satisfying the group property $\Phi_{t+s}=\Phi_t\circ\Phi_s$, $s,t\ge 0$, and such that $t\mapsto \Phi_t$ is continuous, is called a {\it quantum dynamical semigroup}, or a {\it Markovian semigroup} \cite{AlickiLendi}. 
\end{itemize}
The generator of the semigroup $\Phi_t$ is the operator defined by $\ml=\partial_t|_{t=0} e^{t\ml}$ acting on $\mathcal B(\h)$, so that  
\[
\Phi_t = e^{t\ml}.
\]
We now address the following question: What is the general form of an $\ml$ which generates a Markovian semigroup? We will show that such $\ml$ have a specific structure, known as the GKSL form (Gorini-Kossakowski-Sudarshan-Lindblad). To find this form we fix $t$. Being completely positive, $\Phi_t$ has a Kraus representation (generally with time dependent Kraus operators)
\[
\Phi_t(X) = \sum_\alpha K_\alpha(t) XK_\alpha(t)^\dag.
\]
Let $\{F_j\}_{j=1}^{d^2}$ be an orthonormal basis of $\mathcal B(\h)$,
\begin{equation}
	\label{basisF}
\langle F_i,F_j\rangle \equiv \tr (F^\dag_i F_j) = \delta_{ij}
\end{equation}
(this is the inner product of $\mathcal B(\h)$). Choose $F_{d^2}=\frac{1}{\sqrt d}\bbbone$. Then all $F_j$ for $j<d^2$ are traceless, that is, they have trace equal to zero. Expanding in this basis gives, $K_\alpha(t) = \sum_j \langle F_j, K_\alpha(t)\rangle F_j$, and we obtain
\begin{equation}\label{coefficients}
\Phi_t(X) =\sum_{i,j=1}^{d^2} c_{ij}(t) F_i X F^\dag_j,\qquad c_{ij}(t) = \sum_\alpha \langle F_i, K_\alpha(t)\rangle\overline{\langle F_j,K_\alpha(t)\rangle}.
\end{equation}
By the definition $\ml=\partial_t|_{t=0} \Phi_t$ we have for any $X\in\mathcal B(\h)$,
\begin{align}\label{LX}
\ml X &= \lim_{t\rightarrow 0_+}\frac1t \big(\Phi_t(X)-X\big) \nonumber\\
&= \lim_{t\rightarrow 0_+}\Big\{ \sum_{i,j=1}^{d^2-1} \frac{c_{ij}(t)}{t} F_i X F_j^\dag + \frac1d \frac{c_{d^2d^2}(t)-1}{t}X +\frac{1}{\sqrt d}\sum_{i=1}^{d^2-1}\frac{c_{id^2}(t)}{t} F_iX +\frac{1}{\sqrt d}\sum_{j=1}^{d^2-1}\frac{c_{d^2j}(t)}{t} XF_j^\dag\Big\}.
\end{align}

\begin{ex}\label{exercise6}
Define $\Gamma_{ij}(X)=F_iXF_j^\dag$, where $i,j\in\{1,\cdots, d\}$. These are elements of $\mathcal B(\mathcal B(\h))$, linear operators acting on the space of linear operators $\mathcal B(\h)$.  
\begin{itemize}
\item[(a)] Show that the $\{\Gamma_{ij}\}_{i,j=1}^{d^2}$ are $d^2\cdot d^2$ linearly independent elements of $\mathcal B(\mathcal B (\h))$. Consequently, they are a basis of $\mathcal B(\mathcal B(\h))$.
\item[(b)] Show that each term in the sum in \eqref{LX} converges individually as $t\to0_+$. (This fact is often mentioned in textbooks but it is not often proven | the proof is not hugely complicated, but it is a bit delicate). 
\end{itemize}
\end{ex}

We then define for $i,j<d^2$,
\begin{equation}
	\label{Amat}
a_{d^2d^2} =\lim_{t\rightarrow 0_+}\frac{c_{d^2d^2}(t)-d}{t},\quad a_{id^2}=\lim_{t\rightarrow 0_+}\frac{c_{id^2}(t)}{t},\quad a_{d^2j}=\displaystyle \lim_{t\rightarrow 0_+}\frac{c_{d^2j}(t)}{t}, \quad a_{ij}=\lim_{t\rightarrow 0_+}\frac{c_{ij}(t)}{t}
\end{equation}
and we set 
\[
F=\frac{1}{\sqrt d}\sum_{i=1}^{d^2-1} a_{id^2}F_i.
\]
Then 
\[
\ml X = \sum_{i,j=1}^{d^2-1} a_{ij} F_i XF_j^\dag +\frac1d a_{d^2d^2}X +FX+XF^\dag.
\]
Setting ${\rm Re}F=\tfrac12(F+F^\dag)$ and ${\rm Im}F=\tfrac{1}{2i}(F-iF^\dag)$ we get
\[
LX=\sum_{i,j=1}^{d^2-1} a_{ij} F_i X F_j^\dag +\frac1d a_{d^2d^2}X +\big\{{\rm Re}F,X\big\} +i\big[{\rm Im}F,X\big],
\]
where $\{A,B\}=AB+BA$ and $[A,B]=AB-BA$ are the anti-commutator and the commutator, respectively. Now since $\Phi_t$ is trace preserving we have $\tr(\ml X)=0$, which gives
\[
\tr\Big( \sum_{i,j=1}^{d^2-1} a_{ij} F^\dag_jF_i+\frac1d a_{d^2d^2}\bbbone+2{\rm Re}F\Big)X=0\qquad \forall X\in\mathcal B(\h),
\]
which implies that 
\[
{\rm Re}F = -\frac{1}{2d}a_{d^2d^2}\bbbone-\frac12 \sum_{i,j=1}^{d^2-1}a_{ij}F^\dag_j F_i.
\]
Define now 
\[
H=H^\dag=-{\rm Im}F,
\]
then 
\[
\ml X = -i[H,X] +\sum_{i,j=1}^{d^2-1}a_{ij}\big(F_i XF_j^\dag -\tfrac12 \{F_j^\dag F_i,X\}\big).
\]

\begin{ex}\label{exercise7}
    Show that the coefficient matrix $A=(a_{ij})_{i,j=1}^{d^2-1}\, $ is positive definite.
\end{ex}

We diagonalize $A$ by a unitary base change $U$,
\[
A=UDU^\dag,\qquad D={\rm diag}(\gamma_1,\ldots,\gamma_{d^2-1})
\]
where the $\gamma_j\ge 0$ are the eigenvalues. With the definition 
\[
V_l=\sum_{i=1}^{d^2-1} U_{il}F_i
\]
the last formula for $\ml X$ gives the following result.

\begin{thm}[Gorini-Kossakowski-Sudarshan-Lindblad]
\label{thm3}
If a linear operator $\ml$ on $\mathcal B(\mathcal B(\cx^d))$ is the generator of a CPTP semigroup, then there are operators $H=H^\dag$, $V_l\in\mathcal B(\cx^d)$, $l=1,\ldots,d^2-1$ and numbers $\gamma_l\ge 0$ such that 
\[
\ml X = -i[H,X] +\sum_{l=1}^{d^2-1}\gamma_l\Big(V_l X V_l^\dag -\tfrac12 \{ V^\dag_lV_l,X\}\Big).
\]
\end{thm}
One can show that the converse statements holds as well. This theorem and its converse statement is called the {\it Gorini-Kossakowski-Sudarshan-Lindblad} (GKSL) theorem. It was proven by the first three authors for finite dimensional $\h$ in \cite{GKS} and by the last author for bounded $\ml$ in possibly infinite dimensional $\h$ in \cite{Lindblad}. A historical overview of the theorem is given in \cite{ChrPas}.

\section{Solutions to the exercises}

\subsection{Solution to Exercise \ref{ex1}}

In order to calculate the partial trace, we express  $|\phi(t)\rangle\langle\phi(t)|$ as a sum of factorized operators. We simply write $|0\rangle$ and $|1\rangle$ instead of $|0\rangle_\s$ and $|1\rangle_\s$ in the sequel. Using \eqref{wavefunction}, we obtain
\begin{align}
|\phi(t)\rangle\langle\phi(t)|=&|0\rangle\langle0|\Big(|c_0|^2\vac_\r {}_\r\langle{\rm vac}|+c_0\sum_k\overline{d_n(t)}\vac_\r {}_\r\langle k| \\ \nonumber
&+\overline{c_0}\sum_kd_k(t)|k\rangle_\r{}_\r\vacbra+\sum_{kl}d_k(t)\overline{d_l(t)}|k\rangle_\r{}_\r\langle l|\Big) \\ \nonumber
&+|0\rangle\langle1|\Big(c_0\overline{c_1(t)}\vac_\r{}_\r\vacbra+\overline{c_1(t)}\sum_k d_k(t)|k\rangle_\r{}_\r\vacbra\Big) \\ \nonumber
&+|1\rangle\langle0|\Big(c_1(t)\overline{c_0}\vac_\r{}_\r\vacbra+c_1(t)\sum_k \overline{d_k(t)}\vac_\r{}_\r\langle k|\Big) \\ \nonumber
&+|1\rangle\langle1|\Big(|c_1(t)|^2\vac_\r{}_\r\vacbra\Big).
\end{align}
Next we use that $\tr(\vac_\r{}_\r\vacbra)=1$, $\tr(|k\rangle_\r{}_\r\vacbra)=\tr(\vac_\r{}_\r\langle k|)=0$ and $\tr(|k\rangle\langle l|)=\delta_{kl}$ to evaluate the partial trace, 
\[\tr_\r\Big(|\phi(t)\rangle\langle\phi(t)|\Big)=\Big(|c_0|^2+\sum_k|d_k(t)|^2\Big)|0\rangle\langle0|+|c_1(t)|^2|1\rangle\langle1|+c_0\overline{c_1(t)}|0\rangle\langle1|+\overline{c_0}c_1(t)|1\rangle\langle0|.
\]
From $\tr(|\phi(t)\rangle\langle\phi(t)|)=1$ it follows that $|c_0|^2+\sum_k|d_k(t)|^2+|c_1(t)|^2=1$, so that the matrix representation of the atom density matrix in the interaction picture is indeed given by \eqref{5}, in the basis $\{|1\rangle,|0\rangle\}$.

\subsection{Solution to Exercise \ref{ex2}}

In the orthonormal basis $\{|1\rangle,|0\rangle\}$, we have the following matrix representations
\begin{align*}
\sigma_+\sigma_-=
\begin{pmatrix}
1 & 0\\[4pt]
0 & 0
\end{pmatrix}, &&  \sigma_+=
    \begin{pmatrix}
0 & 1\\[4pt]
0 & 0
\end{pmatrix},  && \sigma_-=
    \begin{pmatrix}
0 & 0\\[4pt]
1 & 0
\end{pmatrix}.
    \end{align*}
For the commutator, we obtain
    \[
    [\sigma_+\sigma_-,\rho_{\s,\i}(t)]=\begin{pmatrix}
0 & \overline{c_0}c_1(t)\\[4pt]
-c_0\overline{c_1(t)} & 0
\end{pmatrix}.
    \]
For the other terms,
\begin{align*}
\sigma_-\rho_{\s,\i}(t)\sigma_+=\begin{pmatrix}
0 & 0\\[4pt]
0 & |c_1(t)|^2
\end{pmatrix}, && \{\sigma_+\sigma_-,\rho_{\s,\i}(t)\}= \begin{pmatrix}
2|c_1(t)|^2 & \overline{c_0}c_1(t)\\[4pt]
c_0\overline{c_1(t)} & 0
\end{pmatrix}.
\end{align*}
Using these expressions the right hand side of \eqref{ny2} equals,
    \[
-\tfrac{i}{2}S(t)
    \begin{pmatrix}
0 & \overline{c_0}c_1(t)\\[4pt]
-c_0\overline{c_1(t)} & 0
\end{pmatrix}
+\gamma(t)
\left(
\begin{pmatrix}
0 & 0\\[4pt]
0 & |c_1(t)|^2
\end{pmatrix}
-\frac{1}{2}
\begin{pmatrix}
2|c_1(t)|^2 & \overline{c_0}c_1(t)\\[4pt]
c_0\overline{c_1(t)} & 0
\end{pmatrix}
\right),
    \]
which indeed coincides with the right side of \eqref{ny1}. This show the equality \eqref{ny2}.

\subsection{Solution to Exercise \ref{exercise3}}

\begin{figure}
	\centering
	\includegraphics[width=0.55\linewidth]{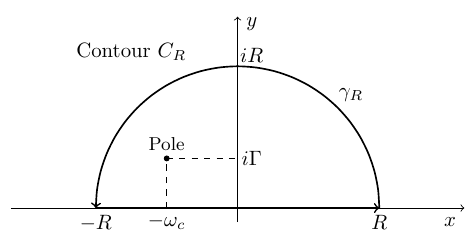}
	\caption{Contour of integration in the complex plane.}
	\label{fig:contour}
\end{figure}

From equation \eqref{ny7},
\begin{equation}
\label{corf}
    f(\tau) =\dfrac{1}{\sqrt{2\pi}}\int_{-\infty}^{\infty}\frac{g}{\pi}\dfrac{\Gamma}{(\omega-\omega_c)^2+\Gamma^2}e^{-i(\omega-\omega_0) \tau}d\omega 
     =\dfrac{g\Gamma e^{i\omega_0\tau}}{\pi\sqrt{2\pi}}\int_{-\infty}^{\infty}\dfrac{e^{ix\tau}}{(x+\omega_c)^2+\Gamma^2}dx,
\end{equation}
where we made the change of variable $x=-\omega$ in the second equality. Write the integrand in the last expression as \[
h(x) = \dfrac{e^{ix\tau}}{(x+\omega_c)^2+\Gamma^2}
\]
and set (for $R>0$)
\[
\int_{-R}^Rh(x)dx=\int_{C_R}h(z)dz-\int_{\gamma_R} h(z)dz,
\]
where $\gamma_R$ is the semicircle of radius $R$ in the upper half plane centered at the origin, and $C_R$ is the union of this semicircle with the horizontal segment from $-R$ to $R$ (see Figure \ref{fig:contour}). 

On $\gamma_\r$ the integration variable is $z=Re^{i\theta}$ with $\theta\in[0,\pi]$. We have 
\begin{align}
\left|\int_{\gamma_R}\dfrac{e^{iz\tau}}{(z+\omega_c)^2+\Gamma^2}dz\right| & \leq \pi R \max_{\theta\in[0,\pi]}\dfrac{e^{-R\tau\sin\theta}}{|(Re^{i\theta}+\omega_c)^2+\Gamma^2|} \le \max_{\theta\in[0,\pi]}\dfrac{\pi R}{\left|(Re^{i\theta}+\omega_c)^2+\Gamma^2\right|},
\label{22}
\end{align}
where we used that $e^{-R\tau\sin\theta}\leq 1$ for $\theta\in[0,\pi]$ in the last step. Next,
\begin{align}
\left|(Re^{i\theta}+\omega_c)^2+\Gamma^2\right|&=\left|R^2e^{2i\theta}+2R \omega_c e^{i\theta}+(\omega_c^2+\Gamma^2)\right| \geq \Big|R|R e^{i\theta}+2\omega_c |-(\omega_c^2+\Gamma^2)\Big|\nonumber \\
&= R^2 \Big| \big|1+\frac{2\omega_c}{R}e^{-i\theta}\big| -\frac{\omega_c^2+\Gamma^2}{R}\Big|.
    \label{RR}
\end{align}
For $R$ sufficiently large, the factor multiplying $R^2$ on the right side of \eqref{RR} is larger than or equal to $1/2$, uniformly in $\theta\in[0,\pi]$.
Therefore we obtain from \eqref{22},
\[
\left|\int_{\gamma_R}\dfrac{e^{iz\tau}}{(z+\omega_c)^2+\Gamma^2}dz\right|\leq \frac{2\pi}{R} \rightarrow 0,\qquad \text{as $R\rightarrow\infty$.}
\]

The function $h(z)$ has a simple pole at  $z_0=-\omega_c+i\Gamma$ inside the contour $C_R$ and so the residue theorem gives, 
\[
\int_{C_R}h(z)dz=2\pi i \text{Res}_{z_0}(h(z))=\frac{\pi}{\Gamma} e^{i(-\omega_c+i\Gamma)\tau}.
\]
Taking $R\to\infty$ we conclude that $ \int_{-\infty}^\infty h(x)dx=\frac{\pi}{\Gamma} e^{i(-\omega_c+i\Gamma)\tau}$ and the correlation function \eqref{corf} becomes
\[
f(\tau)=\frac{g}{\sqrt{2\pi}}e^{-(\Gamma-i\Delta)\tau},\qquad \Delta=\omega_0-\omega_c.
\]
This shows the first part of the exercise. Now we solve the integro-differential equation \eqref{ny3}. Performing the Laplace transform on each side, we obtain
\begin{equation}\label{laplace}
rK(r)-c_1(0)=-F(r)K(r),
\end{equation}
where $K(r)$ and $F(r)$ are the Laplace transforms of $c_1(\tau)$ and $f(\tau)$, respectively. We find
\[
F(r)=\frac{g}{\sqrt{2\pi}}\dfrac{1}{\Gamma-i\Delta+r}.
\]
Solving for $K(r)$ in \eqref{laplace}, we obtain 
\begin{align}
K(r)&=\dfrac{c_1(0)}{r+F(r)}=\dfrac{c_1(0)}{r+\dfrac{g}{\sqrt{2\pi}}\dfrac{1}{\Gamma-i\Delta +r}} =c_1(0)\dfrac{\Gamma-i\Delta +r}{r(\Gamma-i\Delta +r)+\dfrac{g}{\sqrt{2\pi}}} \nonumber \\[9pt] 
    &=c_1(0)\dfrac{\Gamma-i\Delta +r}{\left(r+\dfrac{\Gamma-i\Delta }{2}\right)^2-\dfrac{1}{4}\left((\Gamma-i\Delta)^2-\dfrac{4g}{\sqrt{2\pi}}\right)}. \nonumber
\end{align}
We completed the square in the denominator above for convenience when taking the inverse Laplace transform. 
With the definition \eqref{Rdef} of $R$ we have
\begin{align}
    K(r)=c_1(0)\dfrac{r+\dfrac{\Gamma-i\Delta}{2}}{\left(r+\dfrac{\Gamma-i\Delta }{2}\right)^2-\left(\dfrac{R}{2}\right)^2}+c_1(0)\left(\dfrac{\Gamma-i\Delta}{R}\right)\dfrac{\dfrac{R}{2}}{\left(r+\dfrac{\Gamma-i\Delta }{2}\right)^2-\left(\dfrac{R}{2}\right)^2}.
\end{align}
Using the inverse Laplace transforms
\[
\mathcal L^{-1}\left\{\dfrac{\alpha}{s^2-\alpha^2}\right\}=\sinh (\alpha \tau), \text{ and } \mathcal L^{-1}\left\{\dfrac{s}{s^2-\alpha^2}\right\}=\cosh (\alpha \tau),
\]
we obtain the solution \eqref{solutionDE} to the integro-differential equation \eqref{ny3}.

\subsection{Solution to Exercise \ref{exercise4}}

Set $N=\dim\mathcal{H}$ and let $\{|i\rangle\}_{i=1}^N$ be an orthonormal basis for $\mathcal{H}$. Then
\begin{equation*}
\tr(AX)=\sum_{i=1}^N\langle i|AX|i\rangle =\sum_{i,j=1}^N\langle i|A|j\rangle \langle j|X|i\rangle=\sum_{i,j=1}^NA_{ij}X_{ji}=0,
\end{equation*} 
where $A_{ij}=\langle i|A|j\rangle$ and $X_{ij}=\langle i|X|j\rangle$. Since $X$ is arbitrary, this means that $A_{ij}=0$ $\forall i,j=1,...,N$. Therefore $A=0$.

\subsection{Solution to Exercise \ref{exercise5}}

Let $\{e_1,\cdots, e_d,f_1,\cdots,f_{N-d}\}$ be an orthonormal basis of $\mathcal H$, where $\{e_j\}_{j=1}^d$ is an orthonormal basis of $M$ and $\{f_j\}_{j=1}^{N-d}$ is an orthonormal basis of $M^\perp$. Set $\varepsilon_j=\mathcal Ue_j$ so that $\{\varepsilon_j\}_{j=1}^d$ are $d$ orthonormal vectors in $\mathcal H$, as
\[
\langle\varepsilon_j,\varepsilon_k\rangle=\langle\mathcal U e_j,\mathcal U e_k\rangle=\langle e_j,e_k\rangle=\delta_{jk}.
\]
Complete the set $\{\varepsilon_j\}_{j=1}^d$ to an orthonormal basis
\[
\{\varepsilon_1,\ldots, \varepsilon_d,\eta_1,\ldots,\eta_{N-d}\}
\]
of $\mathcal H$.
Define the operator $\overline{\mathcal U}$ on $\mathcal H$ by setting $\overline{\mathcal U}e_j=\varepsilon_j$ for $j=1,\ldots,d$  and $\overline{\mathcal U}f_k=\eta_k$ for $k=1,\ldots,N-d$.

Since the action of $\overline{\mathcal U}$ on the basis $\{e_j\}_{j=1}^d$ of $M$ is the same as the action of $\mathcal U$, we conclude that $\overline{\mathcal U}$ satisfies $\overline{\mathcal U}|_M=\mathcal U$. Furthermore, the columns $\{\varepsilon_1,\ldots,\varepsilon_d,\eta_1,\ldots,\eta_{N-d}\}$ of the matrix representation of $\overline{\mathcal U}$ are orthogonal, so $\overline{\mathcal U}$ is unitary.

\subsection{Solution to Exercise \ref{exercise6}}

(a) Let $\mathcal U$ be a unitary that changes the ONB $\{F_\alpha\}_{\alpha=1}^{d^2}$ into the ONB $\{E_{ij}\}_{i,j=1}^d$, where $E_{ij}$ is the matrix whose only non-zero entry is $(i,j)$, which is equal to 1. Let $c_{\alpha\beta}$ be complex numbers, $\alpha,\beta=1,\ldots,d$. Then
\begin{equation}
\sum_{\alpha \beta}c_{\alpha\beta} \Gamma_{\alpha\beta}(X)=0  \iff  \sum_{\alpha \beta}c_{\alpha\beta} F_\alpha XF_\beta^\dag=0 
\iff \sum_{\alpha \beta}c_{\alpha\beta} (\mathcal UF_\alpha) X(\mathcal UF_\beta)^\dag=0. \label{esta}
\end{equation}
In order to arrive at the last equivalence, we simply applied $\mathcal U\cdot \mathcal U^\dag$ (for $\Rightarrow$ and $\mathcal U^\dag\cdot \mathcal U$ for $\Leftarrow$). Expanding in the basis $E_{ij}$ we can write 
\begin{equation*}
\mathcal U F_\alpha=\sum_{ij}u_{(ij),\alpha}E_{ij}, \qquad (\mathcal U F_\alpha)^\dag=\sum_{kl}\overline{u_{(kl),\beta}}(E_{kl})^\dag=\sum_{kl}\overline{u_{(kl),\beta}}E_{lk},
\end{equation*}
where the $u_{(ij),\alpha}\in\mathbb C$ are coordinates. Then \eqref{esta} is true if and only if
\begin{equation}
\label{aqui}
\sum_{ij}\sum_{kl}d_{(ij),(kl)}E_{ij}XE_{lk}=0,
\qquad \text{with \qquad 
    $d_{(ij),(kl)}=\sum_{\alpha\beta}c_{\alpha\beta} u_{(ij),\alpha}\overline{u_{(kl),\beta}}$.}
\end{equation}
Now $(\mathcal U^\dag)_{\beta,(kl)}=\overline{(\mathcal U)_{(kl),\beta}}=\overline{u_{(kl),\beta}}$, so
\[d_{(ij),(kl)}=(\mathcal UC\mathcal U^\dag)_{(ij),(kl)}, \quad \text{with\quad $C=(c_{\alpha\beta})$}.
\]
Next, $E_{ij}XE_{lk}=|e_i\rangle\langle e_k|X_{jl}$, where $X_{jl}=\langle e_j, Xe_l\rangle$ is the matrix element of $X$ in the canonical basis $\{|e_j\rangle\}_{j=1}^d$ of $\mathbb C^d$  (and we have $E_{ij}=|e_i\rangle\langle e_j|$).

Choose $X=E_{rs}$, for fixed $r,s$. Then from \eqref{esta} and \eqref{aqui} we have  
\begin{equation*}
\sum_{\alpha\beta}c_{\alpha\beta}\Gamma_{\alpha\beta}(X)=0 \iff \sum_{ik}d_{(ir),(ks)}|e_i\rangle\langle e_k|=0 \iff d_{(ir),(ks)} =0,\quad  \forall i,k=1,\ldots,d.
\end{equation*}
We can choose $r,s$ arbitrary, so 
\begin{equation*}
\sum_{\alpha\beta}c_{\alpha\beta}\Gamma_{\alpha \beta}=0  \iff d_{(ij),(kl)}=0 \quad \forall ijkl \iff\mathcal UC\mathcal U^\dag=0 \iff C=0.
\end{equation*}
This shows that $\sum_{\alpha\beta}c_{\alpha\beta}\Gamma_{\alpha \beta}=0 \Rightarrow c_{\alpha\beta}=0$ for all $\alpha, \beta$. So the  $\Gamma_{\alpha\beta}$ are linearly independent (operators).
\medskip

(b) In terms of the $\Gamma_{ij}$, we have (see \eqref{LX})
\begin{align*}
\ml= \lim_{t\rightarrow 0_+}\Big\{ \sum_{i,j=1}^{d^2-1} \frac{c_{ij}(t)}{t} \Gamma_{ij} +  \frac{c_{d^2d^2}(t)-1}{t}\Gamma_{d^2d^2} +\sum_{i=1}^{d^2-1}\frac{c_{id^2}(t)}{t} \Gamma_{id^2} +\sum_{j=1}^{d^2-1}\frac{c_{d^2j}(t)}{t} \Gamma_{d^2j}\Big\}.
\end{align*}
For short, we write the last relation as 
\[
\ml=\lim_{t\to0_+}\sum_s\xi_s(t)\Gamma_s,
\] 
where $\xi_s(t)\in\mathbb C$ and the $\{\Gamma_s\}$ form a basis of $\mathcal B(\mathcal B(\h))$. $\ml$ has an expansion in the basis, $\ml=\sum_s\lambda_s\Gamma_s$, for some unique $\{\lambda_s\}\subset\mathbb C$. Thus we have 
\[
\lim_{t\to0_+}\sum_s(\xi_s(t)-\lambda_s)\Gamma_s=0.
\]
As the $\Gamma_s$ are linearly independent, there exists $c>0$ such that
\[
\big\|\sum_s(\xi_s(t)-\lambda_s)\Gamma_s\big\|\geq c\sum_s|\xi_s(t)-\lambda_s|.
\]
As the left hand side converges to zero in the limit $t\rightarrow 0_+$ it follows that $\lim_{t\rightarrow 0_+}\xi_s(t)=\lambda_s$ for all $s$.

\subsection{Solution to Exercise \ref{exercise7}}

The entries of the coefficient matrix are defined in \eqref{Amat} by \begin{equation}
	\label{a=c}
a_{ij}=\lim_{t\to0_+}\dfrac{c_{ij}(t)}{t}, \qquad j=1,\ldots,d^2-1.
\end{equation}
We showed in Exercise \ref{exercise6}  that this limit exists.  In view of \eqref{coefficients} we introduce the operator $C(t)=\sum_{\alpha}|K_\alpha(t)\rangle\langle K_\alpha(t)|$, so that $c_{ij}(t)$ is the matrix element of $C(t)$ in the basis $\{F_i\}_{i=1}^{d^2-1}$, \eqref{basisF}. If $C(t)\ge0$ for all $t>0$ then$ \frac{1}{t}C(t)\ge 0$ for all $t>0$ and so by \eqref{a=c} we have $A\ge0$. To show that $C(t)\ge0$ it suffices to prove that $\langle X,C(t) X\rangle\geq 0$ for every $X\in \mathcal B(\mathcal H)$. We have
  \[
    \langle X,C(t) X\rangle=\sum_{\alpha}\langle X|K_\alpha(t)\rangle\langle K_\alpha(t)|X\rangle=\sum_{\alpha}\langle X|K_\alpha(t)\rangle\overline{\langle X|K_\alpha(t)\rangle}=\sum_{\alpha}|\langle X|K_\alpha(t)\rangle|^2\geq 0.
    \]
This concludes the proof.
\bigskip

{\bf Acknowledgements.} We gratefully acknowledge the support of the Natural Sciences and Engineering Research Council of Canada (NSERC) under the Discovery Grant program. MM thanks Graeme Pleasance and Francesco Petruccione for having been invited to deliver these lectures at the 33rd Chris Engelbrecht Summer School on the Theoretical Foundations of Quantum Science, at Stellenbosch University. We are also grateful to two anonymous referees for their constructive feedback on the manuscript.


\begin{thebibliography}{99}

\bibitem{AlickiLendi}
R.~Alicki , K.~Lendi: {\it Quantum Dynamical Semigroups and Applications}, Lecture Notes in Physics, Volume 717, Springer Verlag 2007

\bibitem{hermite}
G. B.~Arfken, H. J.~Weber, and F. E. ~Harris,
{\it Mathematical Methods for Physicists: A Comprehensive Guide},
7th ed., Academic Press, (2012)

\bibitem{Attal}
S.~Attal: {\it Quantum Channels}, https://ncatlab.org/nlab/files/Attal-QuantumChannels.pdf

\bibitem{BohmGadella}
A.~Bohm and M.~Gadella,
\textit{Dirac Kets, Gamow Vectors and Gel'fand Triplets: The Rigged Hilbert Space Formulation of Quantum Mechanics},
Lecture Notes in Physics, vol.~348, Springer-Verlag, Berlin, 1989.

\bibitem{BP}
H.P.~Breuer, F.~Petruccione: {\it The theory of open quantum systems}, Oxford University Press 2002

\bibitem{ChrPas}
D.~Chru\'sci\'nski, S.~Pascazio:  {\it A Brief History of the GKLS Equation}, Open Systems \& Information Dynamics, Vol.~{\bf 24}, No.~3, 1740001 (2017)

\bibitem{Davies}
E.~B. Davies,
\textit{Quantum Theory of Open Systems},
Academic Press, London, 1976.

\bibitem{Garraway}
G.M.~Garraway: {\it Nonperturbative decay of an atomic system in a cavity}, Phys.~Rev.~A {\bf 55}(3), 2290 (1997) 

\bibitem{GKS}
V.~Gorini, A.~Kossakowski, E.C.G.~Sudarshan: {\it 
Completely positive dynamical semigroups of $N$‐level systems}, 
J.~Math.~Phys.~{\bf 17}, 821–825 (1976)

\bibitem{Hall2013}
B. C. Hall,
\textit{Quantum Theory for Mathematicians},
Graduate Texts in Mathematics, vol. 267,
Springer, New York, 2013.
\url{https://doi.org/10.1007/978-1-4614-7116-5}


\bibitem{HHH}
M.~Horodecki, P.~Horodecki, R.~Horodecki: {\it Separability of mixed states: necessary and sufficient conditions}, Phys.~Lett.~A, {\bf 223}, 1 (1996)


\bibitem{LiZouShao}
J.G.~Li , J.~Zou, and B.~Shao: {\it Non-Markovianity of the damped Jaynes-Cummings model with detuning},  Phys.~Rev.~A {\bf 81}, 062124 (2010)

\bibitem{Lindblad}
G.~Lindblad: {\it On the generators of quantum dynamical semigroups}, Commun.~Math.~Phys.~{\bf 48}, 119–130 (1976)

\bibitem{MM20}
M.~ Merkli: {\it Quantum Markovian master equations: Resonance theory shows validity for all time scales},  Ann.~Phys.~{\bf 412}, 16799 (29pp) (2020)  

\bibitem{MM22}
M.~ Merkli: {\it  Dynamics of Open Quantum Systems I, Oscillation and Decay}, Quantum {\bf 6}, 615 (2022) and {\it Dynamics of Open Quantum Systems II, Markovian Approximation}, Quantum {\bf 6}, 616 (2022)    

\bibitem{MM23}
M.~ Merkli: {\it  Correlation decay and Markovianity in open systems},  Ann.~Henri Poincar\'e {\bf 24}, 751–782 (2022) 

\bibitem{MZ23}
M.~Merkli, M.~Zagrodnik: {\it Stability of PPT in equilibrium states}, Can.~J.~Phys.~{\bf 101}(12), 720-727 (2023)

\bibitem{RH}
\'A.~Rivas, S.F.~Huelga: {\it Open Quantum Systems, An Introduction}, Springer Briefs in Physics, Springer Verlag 2012

\bibitem{SK}
B.W.~Shore, P.L.~Knight: {\it The Jaynes-Cummings Model}, Journal of Modern Optics {\bf 40}(7), 1195-1238 (1993).  

\bibitem{Stinespring}
W.F.~Stinespring: {\it Positive functions on $C^*$-algebras}, Proceedings of the American Mathematical Society {\bf 6}, 211-216 (1955)

\bibitem{TrushEtAl}
A.S.~Trushechkin, M.~Merkli, J.D.~Cresser, J.~Anders: {\it Open quantum system dynamics and the mean force Gibbs state},  AVS Quantum Sci.~{\bf 4}, 012301 (2022)  

\end{thebibliography}
\end{document}